\documentclass[twocolumn,aps,showpacs,superscriptaddress,prl]{revtex4-2}
\usepackage{amsmath}
\usepackage{graphicx}
\usepackage{slashed}
\RequirePackage[pagewise,mathlines]{lineno}

\begin{document}
\title{Discovery of higher order quantum electrodynamics effect for the vacuum pair production}%
\author{Wangmei Zha}\email{first@ustc.edu.cn}\affiliation{University of Science and Technology of China, Hefei, China}
\author{Zebo Tang}\email{zbtang@ustc.edu.cn}\affiliation{University of Science and Technology of China, Hefei, China}
\date{\today}%
\begin{abstract}
  The quantum electrodynamics (QED) vacuum behaves like a polarizable medium which leads to novel quantum effect, such as lepton pair production from vacuum in a extreme strong electromagnetic field. This phenomena became known as Schwinger mechanism and have not yet been observed, which probes the unknown nonperturbative and nonlinear regime of QED. With the developments in laser physics, it seems that the field strength could be approached in the foreseeable future, and the interest has grown enormously both in theory and experiment. Currently, on the earth-based experiments, the critical electromagnetic field can only be realized in relativistic heavy ion collisions. However, the duration time for the strong field in heavy-ion collisions is very short, which makes the perturbative theory still effective in contrary to the nonperturbative situation of the Schwinger mechanism. Surely, there should be sizeable and interesting higher order effect due to the large charge carried by ion (coupling constant $Z\alpha \sim $ 0.6), which links the crossover from perturbative to nonperturbative regime of QED. In 1954, Bethe and Maximon studied the higher order QED effect for a similar process --- Bethe-Heitler process and made a prediction of negative correction. However, the higher order effect has been searched without success since then together with the efforts both from experimentalist and theorist, which greatly challenges the existence of higher order correction for vacuum pair production. In this paper, we show that the lowest order QED calculations for lepton pair vacuum production in heavy-ion collisions are about 20$\%$ higher than the combined world-wide data with a seven sigma-level of significance and the corresponding higher order QED results are consistent with data. We claim the discovery of higher order effect for the QED pair production, which settles the dust of previous debates for several decades. The verification of higher order QED effect is a fundamental scientific problem, which is an important milestone towards the nonperturbative and nonlinear regime of QED vacuum.
\end{abstract}
\maketitle

Quantum electrodynamics (QED) is a quantum field theory of the electromagnetic force~\cite{PhysRev.76.790,PhysRev.76.749,PhysRev.76.769}, which describes the gauge invariant interaction of charged particles with photons. Due to the quantum fluctuations, the QED vacuum behaves like a polarizable medium, leading to novel quantum effects, such as Casimir effect~\cite{Casimir:1948dh}, elastic photon-photon scattering~\cite{Aaboud:2017bwk,Aad:2019ock} and pair production from vacuum in extreme strong electromagnetic field~\cite{PhysRev.82.664}.  The vacuum pair production is a process in which virtual dipole pairs in the vacuum can be accelerated apart by the external field, emerging as real pair. The paradigm of vacuum pair production is the Schwinger mechanism: the production of electron-positron pairs in a constant spatially uniform electric field. The pair production rate per volume for Schwinger mechanism is given by~\cite{Dunne:2004nc}: 
\begin{equation}
\label{equation1}
\frac{d^{4}n_{e^{+}e^{-}}}{d^{3}xdt} \sim \frac{c}{4\pi^{3}\lambda^{4}}exp(-\pi\frac{E_{c}}{E}),
\end{equation}
where $E_{c}$ the critical field strength, $m$ is the electron mass, and $\lambda$ is the Compton wavelength of electron. In the equation, there is no power series expansion in terms of the external field strength $E$, which clearly demonstrates the nonperturbative and nonlinear nature of the production. Currently, our understanding of perturbative regime of QED is extreme good, however, very little is known about the nonperturvative regime. To reach a sizeable production rate from Schwinger mechanism, as shown in Eq.~\ref{equation1}, the external field strength should be stronger than the critical value $E_{c}$($\simeq 1.6 \times 10^{16}$ V/cm). In the realm of astrophysics, it is generally believed that the critical field is realized during the process of gravitational collapse leading to a Black Hole~\cite{PhysRevLett.35.463}, creating a dense plasma of electrons, positrons and photons, which results in the phenomenon of Gamma Ray Bursts (GBRs)~\cite{1970Natur.226..135F,Piran:2004ba,RUFFINI20101}.  Back to the Earth-based laboratories, the extreme critical value prevents the experimentalist to directly discovery this fundamental effect for more than 65 years.

\begin{figure}
	\centering
	\includegraphics[width=0.5\textwidth]{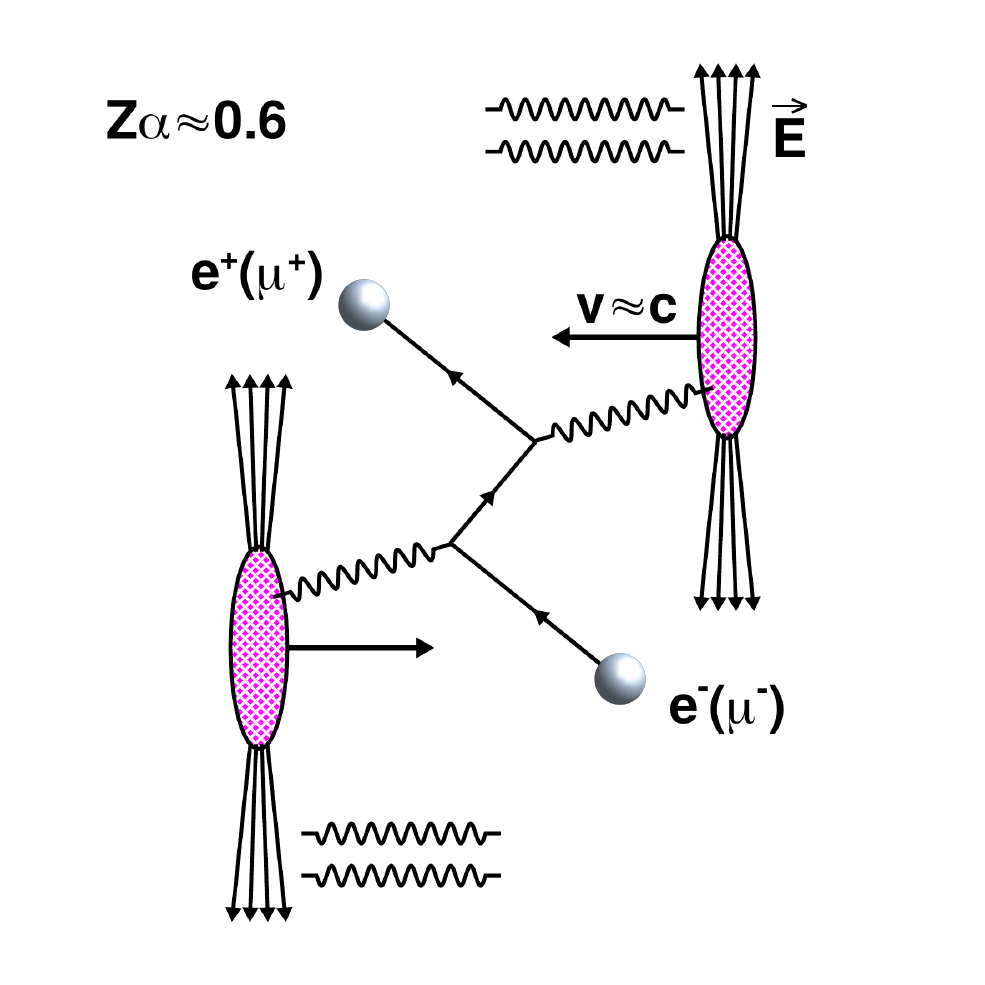}
	\caption{(Color online)  An illustration of vacuum pair production process from the strong Coulomb field of fast moving nuclei in heavy ion collisions.}
	\label{Figure1}
\end{figure}

With the technological advances in laser science~\cite{1960Natur.187..493M,Strickland:1985gxr,Li:18}, it seems that the critical field strength could be realized in the near future. However, the maximum field strength created by the existing PetaWatt class lasers is still orders of magnitude smaller than the critical value. Currently, in the laboratory, the critical field strength can only be reached in the strong Coulumb field  of relativistic heavy ion collisions~\cite{RUFFINI20101}. The electromagnetic fields accompanied in these collisions are of order of~\cite{BAUR20071}
\begin{equation}
\label{equation2}
E_{max} \simeq \frac{Ze\gamma}{b^{2}},
\end{equation}
where $Ze$ is the electric charge carried by the colliding ion, $b$ is the impact parameter, and $\gamma$ is the Lorentz factor of nuclei in center of mass frame. For gold-gold collisions at highest energies from the Relativistic Heavy Ion Collider (RHIC), with the values of $Z = 79$, $b = 15 fm$ and $\gamma = 108$, the maximum field strength is of $E_{max} = 5.3 \times 10^{16}$ V/cm. For the lead-lead collisions at the Large Hadron Collider (LHC), we get $E_{max} = 1.4 \times 10^{18}$ V/cm with the values of $Z = 82$, $b = 15 fm$ and $\gamma = 2706$. The field strengths generated at RHIC and LHC are comparable to the critical field $E_{c}$ for Schwinger mechanism. However, the duration time of the strong field is very short, which makes the perturbative theory still appropriate in contrast to the nonperturbative situation of the Schwinger production. The theoretical investigation of vacuum pair production in heavy-ion collisions goes back to the early days of QED. The lowest order (Born approximation) results were given by Landau and Lifshitz~\cite{Landau} and Racah~\cite{Racah} in the thirties of the past century.  In the calculations, as illustrated in Fig.~\ref{Figure1}, the pair production is treated via the collision of two quasi-real light quanta from the strong Coulumb field surrounding the heavy ions.  

\renewcommand{\floatpagefraction}{0.75}

Due to the large charge carried by the heavy ion, the effect coupling $Z\alpha$ ($\sim$ 0.6 for gold and lead) for the production process is close to 1. This suggests significant higher order effect, which links the crossover from perturbative to nonperturbative regime of QED. This is analogous to the case of Quantum Chromodynamics (QCD)~\cite{PhysRevLett.30.1346,PhysRevLett.30.1343} from high energy to low energy area. In 1954, the pioneer studies~\cite{PhysRev.93.768,PhysRev.93.788} for higher order effect of QED were made by Bethe, Maximon and Davies in a similar process --- Bethe-Heitler process~\cite{PhysRev.93.768} (the photoproduction of electron-positron pairs in the nuclear Coulomb field). Higher order effects were treated using the Sommerfeld-Maue wave functions, which are appropriate solutions of the Dirac equation at high energy. This approach takes higher order effects into account to all orders and can be related to the usual Feynman graph technique~\cite{PhysRevD.57.4025}. A sizeable negative correction was found to the Bethe-Heilter formula. Come back to the case of vacuum pair production in relativistic heavy ion collisions, the correction should be stronger, since the projectile quasi-real photon is also attached to the heavy ion in contrast to the Bethe-Heilter process. The theoretical progresses of higher order QED effects in heavy-ion collisions are reviewed in Ref.~\cite{BAUR20071}.

The experimental investigations of vacuum pair production in heavy ion collisions were spurred in the late 20th century by the relativistic heavy ion facilities Alternating Gradient Synchrotron (AGS) at Brookhaven, the Super Proton Synchrotron (SPS) at the European Organization for Nuclear Research (CERN) and more recently RHIC at Brookhaven and LHC at CERN. Over the decades, various measurements of lepton pair production~\cite{PhysRevLett.69.1911,BAUER1994471,PhysRevC.70.031902,Afanasiev:2009hy,Abbas:2013oua,Sirunyan:2018fhl,Aad:2020dur} has been made in ultra-peripheral collisions (UPC) based on these facilities. Surprisingly, all the measurements are found to be in good agreement with the lowest order calculation. The searching for higher order effect seems to be hopeless, just as the statement in Ref.~\cite{BAUR20071}: “In April 1990 a workshop took place in Brookhaven with the title ’Can RHIC be used to test QED?’. We think that after about 17 years the answer to this question is ’no’.”. This is very puzzling: if the higher order effects are not there, how could we expect the nonperturbative vacuum pair production from Schwinger mechanism? And exploring the nonperturbative feature of QED is one of the primary goal for the planing ultrastrong laser facilities in the not-so-far future.

In this paper, we have performed the latest lowest order QED calculations for lepton pair production in heavy-ion collisions and found that the results are about seven standard deviations larger than the combined world-wide data.  Furthermore, the higher order QED corrections to all orders have been carried out, which reduce the lowest order results sizeably ($~\sim 20\%$). Taken the higher order effect into account, the corresponding results are consistent with data, which claims the discovery of higher order effect for the QED pair production, and settles the dust of previous debates for several decades. We also discuss the missing part in the old lowest order calculations, which prevents us from the observation of higher order QED effect.

The total cross section of vacuum pair production in lowest order due to the Coulomb fields of two colliding nuclei has been carried out for the first time by Landau and Lifshitz~\cite{Landau} in 1934. However, the experimental measurements are usually performed in limited kinematics phase due to the limited acceptance covered by the detector system, which calls for differential theoretical calculations. We treat the electromagnetic fields of the heavy ions as external fields and employ "straight line approximation" since the deflection angle due to the Coulomb scattering is very small. Following the derivation of Refs.~\cite{Hencken:1994my,Alscher:1996mja}, with the direct and cross Feynman diagrams, one can get the differential probability $\frac{d^{6}P(\vec{b})}{d^{3}p_{+}d^{3}p_{-}}$ in lowest order QED for symmetric heavy-ion collisions ($Z = Z_{A} = Z_{B}$)  as
\begin{equation}
\label{equation2_new1}
\frac{d^{6}P(\vec{b})}{d^{3}p_{+}d^{3}p_{-}} = \int d^{2}b \frac{d^{6}\widetilde{P}(\vec{q})}{d^{3}p_{+}d^{3}p_{-}} e^{i {\vec{q}} \cdot  {\vec{b}}},
\end{equation}
where $p_{+}$ and $p_{-}$ are the momenta of the created leptons. The introduced Fourier transformation of differential probability $\frac{d^{6}\widetilde{P}(\vec{q})}{d^{3}p_{+}d^{3}p_{-}}$ is given by
\begin{equation}
\label{equation2_new}
\begin{split}
\frac{d^{6}P(\vec{q})}{d^{3}p_{+}d^{3}p_{-}} & = (Z\alpha)^{4}
\frac{4}{\beta^{2}} \frac{1}{(2\pi)^{6}2\epsilon_{+}2\epsilon_{-}} \int d^{2}q_{1}\\
& F(N_{0})F(N_{1})F(N_{3})F(N_{4})[N_{0}N_{1}N_{3}N_{4}]^{-1} \\
& \times {\rm{Tr}}\{(\slashed{p}_{-}+m)[N_{2D}^{-1}\slashed{u}_{1} (\slashed{p}_{-} - \slashed{q}_{1} + m)\slashed{u}_{2} + \\
& N_{2X}^{-1}\slashed{u}_{2}(\slashed{q}_{1} - \slashed{p}_{+} +m)\slashed{u}_{1}] (\slashed{p}_{+}-m)[N_{5D}^{-1}\slashed{u}_{2}\\
& (\slashed{p}_{-} - \slashed{q}_{1} - \slashed{q} + m)\slashed{u}_{1} + N_{5X}^{-1} \slashed{u}_{1}(\slashed{q}_{1} + \slashed{q} - \slashed{p}_{+} \\
& + m)\slashed{u}_{2}] \},
\end{split}
\end{equation}
with
\begin{equation}
\label{equation3_new}
\begin{split}
& N_{0} = -q_{1}^{2},  N_{1} = -[q_{1} - (p_{+}+p_{-})]^{2},\\
& N_{3} = -(q_{1}+q)^{2}, N_{4} = -[q+(q_{1} - p_{+} - p_{-})]^{2}, \\
& N_{2D} = -(q_{1} - p_{-})^{2} + m^{2},\\
& N_{2X} = -(q_{1} - p_{+})^{2} + m^{2}, \\
&N_{5D} = -(q_{1} + q - p_{-})^{2} + m^{2},\\
& N_{5X} = -(q_{1} + q  - p_{+})^{2} + m^{2},
\end{split}
\end{equation}
where the longitudinal components of $q_{1}$ are given by $q_{10} = \frac{1}{2}[(\epsilon_{+} + \epsilon_{-}) + \beta(p_{+z}+p_{-z})]$, $q_{1z} = q_{10}/ \beta$, $\epsilon_{+}$ and $\epsilon_{-}$ are the energies of the produced leptons, $\beta$ is the velocity of heavy ion, $m$ is the mass of lepton, and the form factor $f(N_{k})$ is Fourier transform of the charge distribution in nucleus. In the calculations, we employ the Woods-Saxon form~\cite{Miller:2007ri} to model the charge distribution of nucleus in spherical coordinates:
\begin{equation}
\rho_{A}(r)=\frac{\rho^{0}}{1+\exp[(r-R_{\rm{WS}})/d]},
\label{equation4}
\end{equation}
where the radius $R_{WS}$ and skin depth $d$ are based on fits to electron-scattering data~\cite{WSD} and $\rho^{0}$ is the normalization factor. The photon propagators $F(N_{k})$ attached to the Coulomb field of heavy ions can be written as
\begin{equation}
\label{equation2_3}
F(N_{k}) = \frac{Z\alpha}{N_{k}} = \frac{Z\alpha}{\frac{w_{k}^{2}}{\gamma^{2}}+p_{Tk}^{2}},\ k = 0, 1, 3, 4,
\end{equation} 
where $Z$ is the charge of nucleus, $\alpha$ ($\sim \frac{1}{137}$) is the fine structure constant, $w_{k}$ and $p_{Tk}$ are the energy and transverse momentum of the photons and  $\gamma$ is the Lorentz contraction factor,

It was pointed out by Ivanov, Schiller, and Serbo~\cite{Ivanov:1998ru} that the higher order effect of lepton pair production in heavy-ion collisions was analogous to the well know Bethe-Heilter process on a heavy target~\cite{PhysRev.93.768}, which shows a negative higher order (Coulomb) correction proportional to $Z^{2}$. These authors went on to calculate the higher order correction to the total cross section by considering higher-order Feynman diagrams in a leading logarithm approximation and also found a significant negative contribution. Lee and Milstein~\cite{Lee:2003fh} later came to essentially the same result for higher order correction by constructing an appropriate regularized transverse integral in the low transverse momentum approximation which analytically solves the higher order correction to all orders. They replaced the photon propagator shown in Eq.~\ref{equation2_3} by a properly regularized expression
\begin{equation}
\label{equation2_4}
F(N_{k}) = \frac{1}{2} \int d\rho \rho J_{0}(p_{Tk}\rho)\{exp[2iZ\alpha K_{0}(\rho \omega_{k}/\gamma)] -1\},
\end{equation}	
where $J_{0}$ and $K_{0}$ are modified Bessel functions. This expression goes over into Eq.~\ref{equation2_3} in the perturbative limit ($Z\alpha \rightarrow 0$). In the calculations, we employ Eq.~\ref{equation2_4} to estimate the higher order corrections.

As revealed in Eq.~\ref{equation2_new}, there are apparent numerical difficulties in evaluating the multiple dimension integration due to the oscillating factor $e^{i {\vec{q}} \cdot  {\vec{b}}}$. A more simple model, EPA approach, is widely used, which is identical to the lowest order QED calculation for the cross section estimation. Since most the experimental results are compared to EPA calculations to test the validity of lowest order QED, we briefly introduce it here. In the approach, the vacuum pair production in heavy-ion collisions can be factorized into a semiclassical and quantum part. The semiclassical part deals with the distribution of quasi-real photons induced by the colliding ions, while the quantum part handles the interactions of photon-photon. It gives:
\begin{equation}
\begin{aligned}
&\sigma (A + A \rightarrow A + A + l^{+}l^{-})
\\
& =\int d\omega_{1}d\omega_{2} n(\omega_{1}) n(\omega_{2}) \sigma(\gamma \gamma \rightarrow l^{+}l^{-}),
\label{equation2}
\end{aligned}
\end{equation}
where $\omega_{1}$ and $\omega_{2}$ are the photon energies from the two colliding beams, and $\sigma(\gamma \gamma \rightarrow l^{+}l^{-})$ is the photon-photon reaction cross-section for lepton pair. The photon flux induced by the heavy ions can be modelled using the Weizs\"acker-Williams method~\cite{KRAUSS1997503}:
\begin{equation}
\label{equation3}
\begin{aligned}
& n(\omega,r) = \frac{4Z^{2}\alpha}{\omega} \bigg | \int \frac{\vec{q}_{\bot}}{(2\pi)^{2}}\vec{q}_{\bot} \frac{f(\vec{q})}{q^{2}} e^{i\vec{q}_{\bot} \cdot \vec{r}} \bigg |^{2}
\\
& \vec{q} = (\vec{q}_{\bot},\frac{\omega}{\gamma})
\end{aligned}
\end{equation}
where $n(\omega,r)$ is the flux of photons with energy $\omega$ at distant $r$ from the center of nucleus. For the point-like charge distribution, the photon flux is given by the simple formula
\begin{equation}
n(\omega,r) = \frac{Z^{2}\alpha}{\pi^{2}\omega r^{2}}x^{2}K_{1}^{2}(x), x = \omega r / \gamma.
\label{equation2}
\end{equation}
The elementary cross-section to produce a lepton pair with lepton mass $m$ and pair invariant mass $W$ can be determined by the Breit-Wheeler formula~\cite{Breit:1934zz}.
\begin{equation}
\label{equation5}
\begin{aligned}
& \sigma (\gamma \gamma \rightarrow l^{+}l^{-}) =
\\
&\frac{4\pi \alpha^{2}}{W^{2}} [(2+\frac{8m^{2}}{W^{2}} - \frac{16m^{4}}{W^{4}})\text{ln}(\frac{W+\sqrt{W^{2}-4m^{2}}}{2m})
\\
& -\sqrt{1-\frac{4m^{2}}{W^{2}}}(1+\frac{4m^{2}}{W^{2}})].
\end{aligned}
\end{equation}

\renewcommand{\floatpagefraction}{0.75}
\begin{figure}
	\centering
	\includegraphics[width=0.5\textwidth]{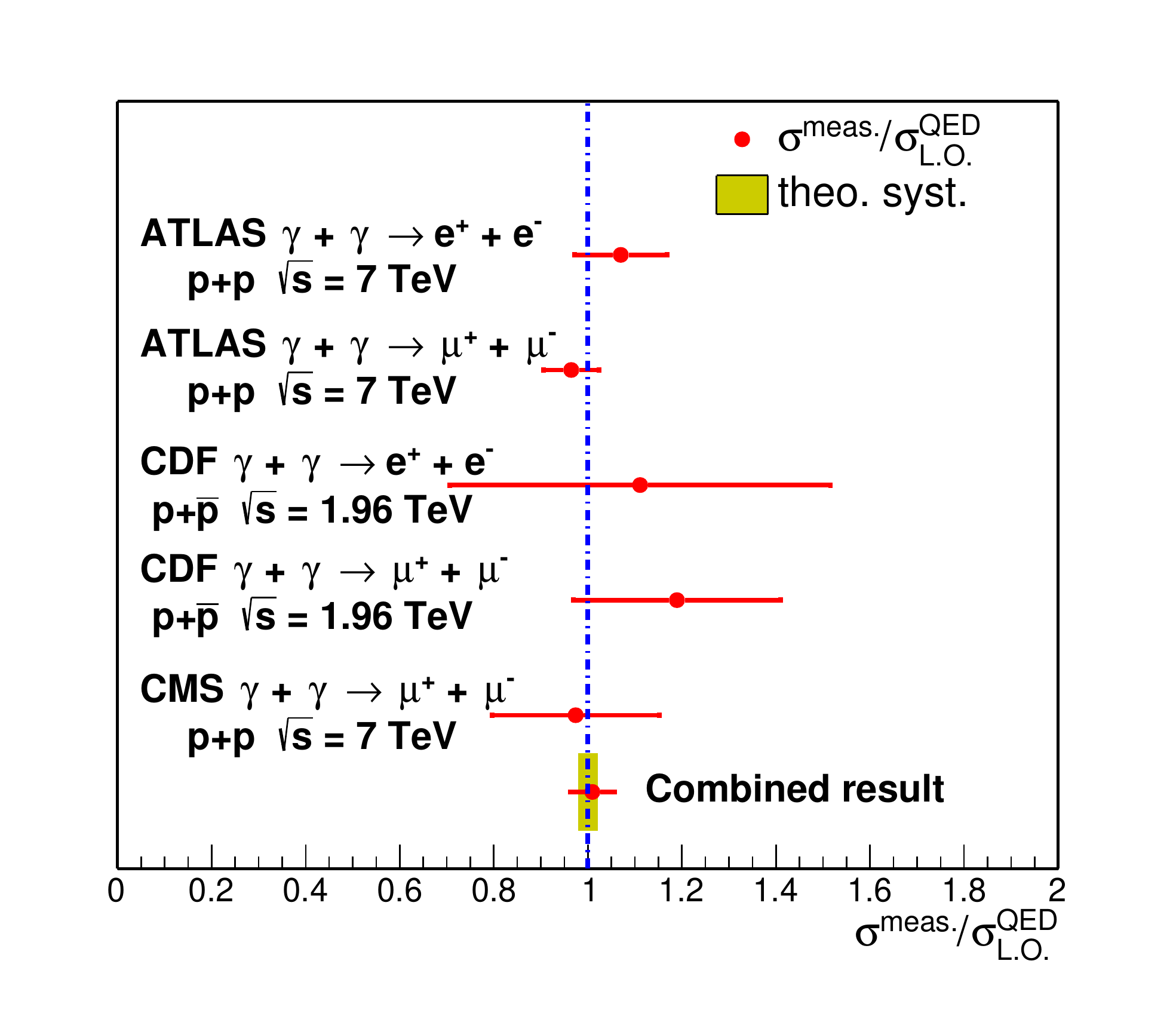}
	\caption{(Color online) Comparison of the ratios of the measured cross sections from world-wide experiments~\cite{Aaltonen:2009kg,Abulencia:2006nb,Aad:2015bwa,Chatrchyan:2012tv} to the lowest order QED calculations for lepton pair production in $p+p(\bar{p})$ collisions. The error bar represents the total uncertainty for each measurement, which includes the statistical and systematic errors. The yellow band denotes the uncertainty for lowest order QED calculations. }
	\label{Figure2}
\end{figure}

Various exclusive experimental measurements~\cite{Aaltonen:2009kg,Abulencia:2006nb,Aad:2015bwa,Chatrchyan:2012tv} for lepton pair production in $p+p(\bar{p})$ have been made, despite of the tiny cross section due to $Z =1$. The coupling constant in these collisions is in perturbative limit ($Z\alpha \rightarrow 0$), which provides excellent baseline to test the validity of lowest order QED. Fig.~\ref{Figure2} shows comparison of the ratios of the measured cross sections from world-wide experiments~\cite{Aaltonen:2009kg,Abulencia:2006nb,Aad:2015bwa,Chatrchyan:2012tv} to the lowest order QED calculations for lepton pair production in $p+p(\bar{p})$ collisions.  The error bar represents the total uncertainty for each measurement, in which the statistical and systematic errors are added in quadrature. The exclusive measurements are performed in UPC events, which exclude the inelastic collisions to reject hadronic background. In the calculations, the UPC trigger probability in impact parameter space can be modelled by
\begin{equation}
\label{equation3_1}
P_{\rm{non-inelastic}} = |1-\rm{exp}(-b^{2}/(2B))|^{2},
\end{equation} 
where $B$ is determined by the experimental measurements~\cite{Zyla:2020zbs,Aad:2011eu}. The standard dipole form factor of proton (anti-proton) is utilized for the lowest order QED calculation. The theoretical uncertainties are estimated by varying the parameters in the proton form factor and the UPC trigger probability, which are found to be less than $2\%$. The yellow band in the figure represents the uncertainty for lowest order QED calculations. As shown in figure, the lowest order QED calculations describe the world-wide measurements very well. The world-wide results are combined with unequal weights determined by the errors to improve the precision of measurement. The combined result is consistent with the lowest order QED calculation within one standard deviation.

\renewcommand{\floatpagefraction}{0.75}
\begin{figure}
	\centering
	\includegraphics[width=0.5\textwidth]{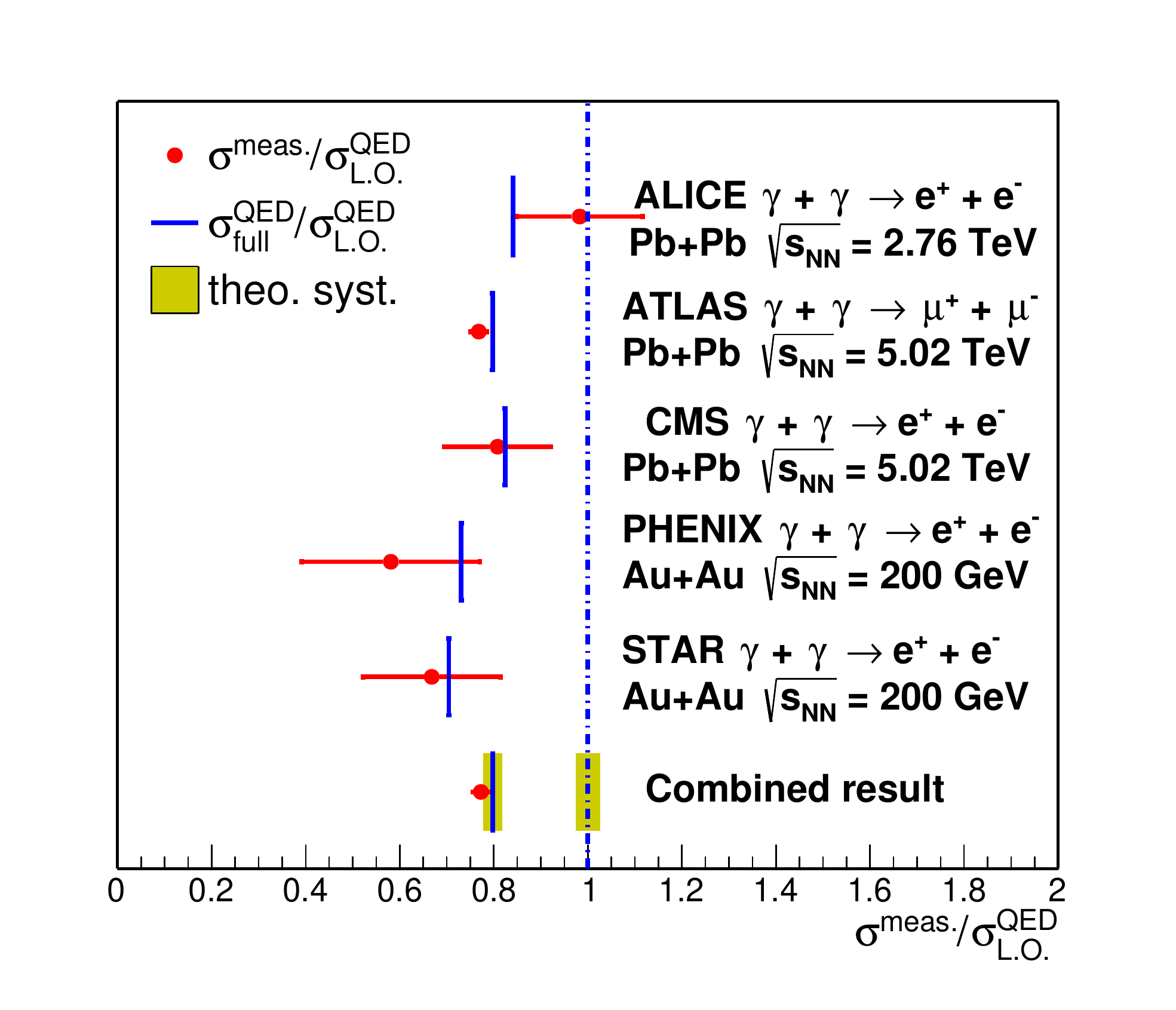}
	\caption{(Color online)  Comparison of the ratios of the measured cross sections from world-wide experiments~\cite{PhysRevC.70.031902,Afanasiev:2009hy,Abbas:2013oua,Sirunyan:2018fhl,Aad:2020dur} and the predicted higher order QED results to the lowest order QED calculations for lepton pair production in $A+A$ collisions. The error bar represents the total uncertainty for each measurement, which includes the statistical and systematic errors. The yellow bands denote the uncertainties for QED calculations.}
	\label{Figure3}
\end{figure}

The comparison of the ratios of the measured cross sections from world-wide experiments~\cite{PhysRevC.70.031902,Afanasiev:2009hy,Abbas:2013oua,Sirunyan:2018fhl,Aad:2020dur} and the predicted higher order QED results to the lowest order QED calculations for lepton pair production in heavy ion collisions is shown in Fig.~\ref{Figure3}. In the calculation, optical Glauber model~\cite{Miller:2007ri} is employed to determine the UPC trigger probability of heavy ion collision in impact parameter space. The theoretical uncertainties are estimated by varying the parameters in the Woods-Saxon distributions of heavy nuclei, which simultaneously changes the form factor of heavy nuclei and the UPC trigger probability in impact parameter space. The uncertainties are found to be less than $2.5\%$, which are represented as yellow bands in the figure. As demonstrated in the figure, the measurements are systematically smaller than the lowest order QED predictions and the QED results with higher order correction describe the data very well. The combined data is seven standard deviations smaller than the lowest order calculation and consistent with the higher order result within one standard deviation, which claims the discovery of higher order effect for the QED pair production in heavy ion collisions.

As shown in Fig.~\ref{Figure3}, the higher order corrections are significant ($~\sim 20\%$). This raises a puzzle: why the higher order effect is not observed previously? Most the experimental measurements are compared to a industry standard model --- STARLight~\cite{Klein:2016yzr}, which incorporates the equivalent photon approximation (EPA) approach to calculate the lepton pair production. The EPA approach is identical to the lowest order QED calculation for cross section estimation. In STARLight model, it treats the nucleus as a point-like charge for evaluating the photon flux. To avoid the singularities in the spatial distribution of photon flux, the pair production within the geometrical radius of the nucleus is ignored. However, according to our study, the production within the geometrical radius of the nucleus is not negligible. As an illustration, for the dimuon production in $Pb+Pb$ collisions at $\sqrt{s_{NN}}$ = 5.02 TeV with the fiducial acceptance of ATLAS ($p_{T\mu} > 5$ GeV/c, $|\eta_{\mu}| <2.4$), the fraction of production within the nuclei is about $28\%$. Coincidently, the higher order correction to the lowest QED result is about $20\%$, which is comparable to the fraction of production ignored in STARLight model. The two missing parts (the production within the nuclei and the higher order correction) compensate with each other, which makes the STARLight model effective to describe world-wide data. This prevents us from the observation of higher order effect.

We report the lowest-order QED calculations for lepton pair production both in proton-proton (anti-proton) and heavy ion collisions. The lowest-order predictions describe the world-wide measurements in $p+p(\bar{p})$ collisions ($Z\alpha \rightarrow 0$) very well, however, overestimate the production in heavy ion collision ($Z\alpha \sim 0.6$) by about $20\%$ with a seven sigma-level of significance. The corresponding higher order QED results can reproduce the world-wide measurements within one standard deviation. These findings lend credence to claim the discovery of higher order effect for the QED pair production under the strongest electromagnetic field in laboratory, which have waited more than half a century for verification and pave the way for future tests of QED in the unexplored nonperturbative regimes. Furthermore, by colliding different species of nuclei, the coupling constant for the QED pair production can be varied to investigate the higher order effect towards the nonperturbative regime, which provides a nice reference for the study of QCD from perturbative to nonperturbative area. The investigations towards nonperturbative QED cover complementary aspects, which shed new light on the understanding of the nuclear and laser physics processes, of heavy ion collisions as well as neutron stars formation and gravitational collapse, supernovae and GRBs phenomena.

We thank Prof. Zhangbu Xu, Prof. Lijuan Ruan, Prof. Spencer Klein and Dr. Daniel Brandenburg for useful discussions. This work was funded by the National Natural Science Foundation of China under Grant Nos. 11775213 and 11720101001, and MOST under Grant No. 2018YFE0104900.

\nocite{*}
\bibliography{aps}
\end{document}